# Is There Really "Retrocausation" in Time-Symmetric Approaches to Quantum Mechanics?

Ruth E. Kastner[1, a)]

[1]*Department of Philosophy, University of Maryland, College Park, MD 20742*
[a)]*Corresponding author: rkastner@umd.edu*

**Abstract.** Time-symmetric interpretations of quantum theory are often presented as featuring "retrocausal" effects in addition to the usual forward notion of causation. This paper examines the ontological implications of certain time-symmetric theories, and finds that no dynamical notion of causation applies to them, either forward or backward. It is concluded that such theories actually describe a static picture, in which the notion of causation is relegated to a descriptor of static relationships among events. In addition, these theories lead to an epistemic rather than ontologically referring, realist view of quantum states.

## INTRODUCTION

Several researchers have proposed interpretations of quantum theory that involves explicit times symmetry [1 – 5]. This paper discusses the inconsistency between causal dynamism and the block world ontology implicit in these time-symmetric interpretations of quantum mechanics. While the present author also has proposed an interpretation involving time-symmetric field propagation based on the direct-action theory of fields (the "possibilist transactional interpretation PTI), [6], my approach has a temporal asymmetry owing to the response of absorbers, which break the time symmetry. (For a discussion of this, see [7],[8].) In addition, PTI treats space-time events as unidirectionally emergent from the underlying time-symmetric processes, which are seen as taking place in a pre-spatiotemporal domain of Heisenbergian *potentiae*. So the observations herein do not apply to PTI, which has a growing space-time ontology in which the future is open; there is no future spacetime boundary condition. The focus of this paper is solely on those time-symmetric theories that seek to 'restore time symmetry' at the level of spacetime, such that the distinction between past and future is merely perspectival rather than ontological. Such interpretations all have future spacetime boundary conditions.

It should be emphasized at the outset that this author is certainly sympathetic towards the philosophical project of exploring time symmetry and investigating whether our experience of temporal directionality is ontologically based or merely perspectival. As far as I can tell, there is no way to settle this question empirically; we have to invoke other considerations such as fruitfulness of the approach, economy and consistency of explanation, realism versus non-realism, etc., in order to try to assess which interpretation might be a correct description of nature. The present paper aims only to observe that what is often said about these theories in dynamical terms cannot really be upheld, since their explicit time symmetry forces a static ontology. Thus my criticism is not based on *a priori* rejection of the block world ontology; it is possible that we do live in a block world. Rather, I simply offer an observation that the dynamical stories presented in these interpretations are inconsistent with the ontology they require.

## CAUSATION, FORWARD AND/OR BACKWARD

Time-symmetric interpretations such as the two-state vector formalism (TSVF) [5] and time symmetric hidden variables (TSHV) such as those of Price and Wharton[1,2] and Sutherland [3,4] are presented as involving a bidirectional causal flow. (We might consider these theories, rather than mere interpretations, of quantum theory, since they all involve additional theoretical structure beyond that of standard quantum mechanics, so I will refer to them as theories in this discussion). In TSVF, a system is taken as being described by both its pre-selection state $|\psi\rangle$ and post-selection state $|\phi\rangle$, through the putatively bi-directional 'two-state vector' $\langle\phi||\psi\rangle$. The theories of Price, Wharton and Sutherland involve time symmetric hidden variables (TSHV); i.e., hidden variables that have dependence on both past and future events. Sutherland's is a time-symmetric version of the Bohmian theory in which position is the hidden variable. Price, in [1], coined the term 'advanced action' for the idea that dynamical influences can propagate from an event E occurring at $t$ toward past times $t_p < t$.

This author is aware of the highly nontrivial task of defining causation, and this paper by no means pretends to offer a categorical definition. I will simply lay out what I think is a common understanding of what it means to say that one event causes or influences another event (the latter being perhaps a weaker proposal but still involving dynamism). I believe this common understanding is what is in play when the above theories are presented, so that is why I think it should suffice for the present discussion. For example, Sutherland expresses the view that retrocausation can 'save locality' in quantum theory, by explaining the distant connections between entangled EPR pairs in terms of a dynamical "zig-zag" influence going backward and then forward again in time between one particle, back to the source, and then forward to the other particle.[1]

In addition, advocates of the TSVF suggest that a 'future event influences a past event' (e.g.,[9]). As noted above, TSVF a system is described by a "two state vector" (TSV), $\langle\phi||\psi\rangle$. Aharonov *et al* explicitly state that the post-selection state propagates toward the past; the TSV is taken as a bi-directional entity. So it is assumed that dynamical propagation is occurring in spacetime. All these time-symmetric theories have in common a final boundary condition on the universe, and a determinate set of spacetime events in between the initial and final boundary condition. (It should be noted that the correlations discussed in [9] are straightforwardly predicted by standard quantum mechanics and therefore do not require explicit spacetime retrocausation.[2])

Let us now see whether we can make sense of the dynamical stories accompanying these theories. First, consider a sequence of events (even a deterministic one) appearing one after another in a single temporal direction (either forward or backward). We can apply a counting process to each sequence: the first event to appear is indexed by 0, the second by 1, etc. If we like, we can index a past-direct sequence by 0, -1, -2, .... ; or we can use the same positive indices and consider them values of -$t$. This is just an arbitrary convention, given a unidirectional sequence of events: an observer experiencing a sequence of events growing in this way, in either direction, would presumably not be aware of any difference in temporal orientation. One event is simply experienced as following another. In this circumstance, one can meaningfully think of event $n$ as the cause of event $n+1$, since the occurrence of event $n$ was a prerequisite for the appearance of event $n+1$. (For the past-directed sequence, one can think of event $n$ as the cause of event $n-1$; event n is a prerequisite for the occurrence of event $n-1$.) Another way to describe the situation is that one event is being *generated* by another event; without its generating event $n$, event $n+1$ would never appear (and *mutatis mutandis* for the past-directed sequence). This account identifies a cause $n$ of an event $n\pm1$ as a necessary and sufficient condition for the occurrence of $n\pm1$.

Thus, there is nothing strange about 'retrocausation' if it is taken as applying to a unidirectional sequence of events. It is just a convention in which the temporal index has the opposite sign. However, it is only in such a unidirectional sequence of events, appearing one after the other (where 'after' is relative to the orientation of the sequence numbering), that one can think of one event as causing another, in this 'generating' sense; the set of events is always increasing. In contrast, in a block world, all the events are given as an entire set, not as a growing sequence. It is not the case that any events are 'coming into being'; all spacetime events exist up to and including the final instant of the universe. Therefore, no event is a prerequisite for, or generator of, any other event, since all events appear with equal priority. Put differently, for any event E occurring at t=0, events D and F preceding and following E have nothing to contribute to E's existence. If event E is King Arthur's coronation, event D is young Arthur pulling the sword from the stone, and F is Merlin approaching from the future to guide the young Arthur, event E has no need for either D or F, since it already exists in spacetime. Neither D nor F is a necessary and/or sufficient condition for E. The only dynamical role that could be played by D and F would be in establishing the spacetime manifold from some orthogonal manifold (which we could think of as 'God creating the block world').

---

[1] This was stated, for example, in Sutherland's presentation at the AAAS Workshop on Retrocausation (San Diego, June 15-16, 2016). The "zig-zag" account of EPR correlations is also given in Cramer's original presentation of the transactional interpretation [10]. This author differs with Cramer in his assumption that all processes are spacetime processes, for that makes his original version of TI subject to the same criticism elaborated herein.

[2] This issue will be elaborated in a separate work.

One might think that since the laws of physics, including quantum mechanics, have a dynamical expression--e.g., the Schrödinger evolution describes a quantum state changing with time--that these laws are 'in play' even in a block world picture. So suppose we think of the spacetime manifold as analogous to the field between capacitor plates, where the plates play the part of the initial and final universal boundary conditions. The establishment of such a field is certainly a dynamical process (charging up the plates); but this corresponds to the initial *establishment* of the boundary conditions and the field (God creating the spacetime manifold, if you will). Once the plates are fully charged and the field established, we have a static situation--an electrostatic field. The latter is what corresponds to the spacetime manifold in the block world of the time-symmetric theories, with their specified boundary conditions and physical laws determining the configuration of all the events. There is no provision in these theories for the 'becoming' stage of the spacetime manifold. Perhaps that could be added as an additional ontological component; but it is not part of the proposals in their current form.

## RETROCAUSATION ABSENT IN BLOCK WORLD TIME-SYMMETRIC THEORIES

According to the above, the block world ontology is static and acausal. This is also the conclusion of proponents of the Relational Block World (RBW) interpretation of quantum mechanics [11], who present their interpretation as an adynamical and acausal one, and who therefore propose a block world interpretation with a consistent ontology. In contrast, there is an ontological inconsistency in TSVF and TSHV theories to the extent that retrocausation is claimed to play an explanatory role in service of such perceived goals as 'saving locality'. For example, Price has also argued that TSHV constitute a local explanation for the Bell correlations. The idea is that each of the particles in an entangled Bell state possesses time-symmetric hidden variables that gives it a dependence not only on its prepared state but on its detection state, so that it is purportedly influenced by both of these. But in fact, all spacetime events are determinate; each particle's entire trajectory already exists in the block world. So there is no influence propagating "anywhen" in spacetime; it is just a story tacked on to a set of events that already exist.

Thus, once we have all events specified in this way, there is no need for any additional story about propagating influences such as 'advanced action'; it is superfluous. In fact it is the static block world that doing the work of 'saving locality', not any dynamical process. Faster-than-light influences are eliminated because the quantum correlations are explained through violation of Einstein's 'being thus' criterion (also called "separability," quantified by Shimony [12] as "outcome independence"), rather than through faster-than-light signaling (quantified by Shimony as "parameter independence"). The point here is that the dynamical story is not an accurate reflection of the ontology of these theories. The different specific theories simply amount to different ways of accounting for the pattern of static relationships among spacetime events.

In addition, in these theories the quantum state has been relegated to an epistemic role: it does not refer to an objective uncertainty but only an epistemic uncertainty. In the view of this author, the epistemic approach is therefore a step toward non-realism about the quantum state, since any statistical uncertainty obtaining in the state descriptions refers only to the ignorance of an observer. Thus, the quantum state description does not ontologically refer. This is certainly an option one can take. However, it appears that proponents of these interpretations see them as realist approaches, which would seem to be in conflict with the fact that their quantum states do not refer to something in the world. One might argue against this that classical statistical mechanics (CSM) could be regarded as realist, so why shouldn't taking the quantum state as a statistical description be considered realist? The difference is that CSM adopts the statistical approach *despite* the fact that in any given situation, the system's properties can be considered determinate; they are in principle precisely specifiable. Thus, in classical physics, there are two levels of description, precise vs. statistical, that have different functions. In CSM, we can choose the statistical level of description by voluntarily introducing an uncertainty allowing the inclusion of other possible states of the system besides the one currently possessed by it. In contrast, there is only a quantum single description -- the quantum state -- which has an irreducible uncertainty not introduced voluntarily by the observer. It is a *choice*, not a necessity, to view that irreducible uncertainty as describing the observer rather than the system itself. Taking a theoretical object as referring to an observer's knowledge rather than to the system clearly constitutes a step away from realism about the quantum state.

It thus seems to this author that the above time-symmetric theories and interpretations of quantum mechanics are presented in a way that does not reflect their ontological nature, in two chief ways: (1) They are presented as dynamical accounts when they are actually adynamical; and (2) They are presented as full-blown realist accounts when they are really epistemic about quantum states. In these theories, quantum states--whether single states or TSVs--are just observer-dependent labels placed on sets of events that are ontologically determinate and which have certain static relations amongst one another in the block world. What is doing the heavy lifting in these theories is the block world. The dynamical stories presented along with these accounts do no actual explanatory work, since there is no real dynamics.

## FREE WILL CONSIDERATIONS

Finally, a few remarks about free will. While this is not a necessary part of my critique, it should be briefly addressed here, since proponents of TSHV theories have claimed that their theories are compatible with free will.

It is typically assumed that in order to have free will, we must (at least) have some causal control over our actions. According to the above arguments, one would therefore not have free will in either the TSVF or TSHV ontology. Suppose one attempted to counter this negative conclusion through the application of particular theory of free will, such as the 'agent interventionist' theory. According to this theory, an agent Alice could be said to have free will if she has a 'control knob' that she can turn, which influences some other event. If her influence is forward-propagating, then she is considered as having free will towards the future by virtue of her control; if her influence is past-propagating, then she is considered as having free will toward the past. In the time-symmetric story told with these theories, theoretically Alice would have free will in both directions; but (as a time-asymmetric creature "moving through the block world' in a particular direction), she has knowledge of past events, so she could only be considered to have free will towards the future because of her ignorance of future events. (The notion that observers 'move through' the block world is taken as primitive in block world theories. It cannot be explained from within the theory and must be assumed as an *ad hoc* principle.[3])

However this ignorance-based account will not do, because (as observed above), Alice actually exerts no dynamical influence in either temporal direction. Moreover, it is not up to her what her knob connects to in her future--that is a uniquely determined event. Her knob is not a control knob; it is just one of a pair of related events. So if Alice thinks she has control, she is simply mistaken. Ignorance of future events does not constitute control over future events, and being mistaken about whether one has control does not constitute free will. This account therefore purports to "save" free will by attributing it to what amounts to a delusion on Alice's part. Incorrect information about events cannot reasonably be taken as any form of control over events. If anything, it signals lack of control: if "information is power," then surely wrong information is (at best) lack of power.

## CONCLUSION

It has been argued that time-symmetric theories and interpretations of quantum mechanics that imply a block world ontology cannot consistently be portrayed as involving dynamical influences propagating within spacetime. Such dynamical stories are just narrative overlays on a static ontology. The block world ontology is what 'defangs' the nonlocal correlations between entangled EPR particles by taking all outcomes as determinate in the block world, so that neither particle has to make a seemingly miraculous instantaneous decision as to which way to spin based on its partner's distant measurement result. The story of the dynamical 'zig-zag' influence from one detector back in time to the source and forward again in time to the other detector is not a description of something that really occurs in the ontology. It is therefore misleading and, in the view of this author, should be dropped.

In addition, these theories relegate quantum states to observer-dependent epistemic descriptions whose statistical uncertainties reflect only the ignorance of observers regarding events that are determinate in the spacetime manifold. Underlying every state attribution is a complete set of determinate spacetime events, about which the observer who assigns the state is ignorant; the state therefore amounts to an epistemic probability distribution. Thus these approaches appear not to be fully realist, in the sense that they do not take the quantum state as referring to an observer-independent entity.

---

[3] It is often supposed that one would be mistaken in thinking that this aspect of experience is something that a physical theory should be expected to explain, thus elevating a shortcoming of a particular theoretical approach to an ostensibly required and right-thinking methodological and metaphysical principle. But in fact, other approaches, such as a growing universe picture, can explain this temporally oriented aspect of our experience. An example is the causal set account of Sorkin et al (e.g., [13]), which is fully compatible with relativity.)

Finally, a general observation: it appears that much of the discussion of time symmetric theories -- both by proponents of such theories and even by more or less neutral researchers -- is being undertaken against a backdrop of declining to take the static block world ontology of the theories seriously. For example, a recent paper by Leifer and Pusey [14] attributes backward-propagating physical influences to the term "retrocausation" but then actually defines it in terms of (epistemic) conditional statistical dependence on future events. Clearly, an epistemic statistical dependence and a propagating physical influence are two completely different things. In foundational studies of quantum theory, the distinction between an epistemic quantity and an ontological, physical process or property is a crucial one that bears on which sorts of interpretations are viable and which are not. Yet very commonly in the literature on-time symmetric theories, epistemic and ontological quantities seem to be viewed as interchangeable. This is a curious situation.

Now, one might reply that nobody really knows what causation is, or even if it exists at all, and that all causation/dynamism talk is really (or is reducible to) talk about our epistemic perspectives as observers. That is, a demand to refrain from applying dynamical narratives and claims to a block world ontology is often met with the claim that such a demand is unreasonable: that nobody should be expected to take the underlying ontology seriously when it comes to talking about physical influences and causation. However, such a response assumes that the restrictions of the static block world ontology apply equally to all possible ontologies, when that is not the case. Both the growing causet model of Sorkin et al, and the poset model of Knuth and colleagues (e.g., [15]), have physical dynamism built into the growth of the set, such that the dynamical experience of an observer corresponds to features of the ontology itself, not to the perspective of observer assumed to be (somehow) moving through a static ontology. In such growing universe ("becoming") models, the future is ontologically open and, on that basis, genuine (non-illusory) intervention by an agent is at least in principle possible.[4] Dismissing these models as "speculative" is not a legitimate response: it could be said that time-symmetric hidden variables models are also "speculative." Thus, we have to avoid a double standard here. If a block world ontology is not strictly required by physical theory (as has been shown by Sorkin [17]), then the methodological claim that all "causation" talk is always necessarily perspectival-only is refuted. In fact, it is in principle possible to have a model in which there is genuine dynamics and even genuine intervention by an agent -- not just the illusion of intervention. Thus, intervention and dynamism are not reducible or equivalent to an epistemic-only account. Time-symmetric hidden variables models (as well as TSVF) cannot have it both ways. They are static, not dynamical, ontologies.

## ACKNOWLEDGMENTS

The author is pleased to thank the organizer, Daniel Sheehan, and participants of the Workshop on Retrocausation (AAAS Conference, San Diego, CA; June 15-16, 2016) for the opportunity to present and discuss this paper and related ideas.

---

[4] Arguments that no real intervention (free will) is possible even with quantum indeterminism are rebutted in [16].